\documentstyle[multicol,aps,psfig,bbox]{revtex} 
\begin{document} 
\newcommand{\vk}{{\vec k}} 
\newcommand{\vK}{{\vec K}}  
\newcommand{\vb}{{\vec b}}  
\newcommand{{\vp}}{{\vecp}}  
\newcommand{{\vq}}{{\vec q}}  
\newcommand{\vQ}{{\vec Q}} 
\newcommand{\vx}{{\vec x}} 
\newcommand{\vh}{{\hat{v}}} 
\newcommand{\tr}{{{\rm Tr}}}  
\newcommand{\beq}{\begin{equation}} 
\newcommand{\eeq}[1]{\label{#1} \end{equation}}  
\newcommand{\half}{{\textstyle\frac{1}{2}}}  
\newcommand{\gton}{\stackrel{>}{\sim}} 
\newcommand{\lton}{\mathrel{\lower.9ex \hbox{$\stackrel{\displaystyle 
<}{\sim}$}}}  
\newcommand{\ee}{\end{equation}} 
\newcommand{\ben}{\begin{enumerate}}  
\newcommand{\een}{\end{enumerate}} 
\newcommand{\bit}{\begin{itemize}}  
\newcommand{\eit}{\end{itemize}} 
\newcommand{\bc}{\begin{center}}  
\newcommand{\ec}{\end{center}} 
\newcommand{\bea}{\begin{eqnarray}}  
\newcommand{\eea}{\end{eqnarray}} 
\newcommand{\beqar}{\begin{eqnarray}}  
\newcommand{\eeqar}[1]{\label{#1} 
\end{eqnarray}}  
 
\title{High $p_T$ Azimuthal Asymmetry  
in Non-central A+A at RHIC}

\author{Miklos~Gyulassy$^{1,2}$,  Ivan~Vitev$^{1}$ and Xin-Nian~Wang$^{2}$ } 
 
\address{$^1$  Dept. Physics, Columbia University,  
       538 W 120-th Street, New York, NY 10027\\ 
      $^2$ Nuclear Science Division, Lawrence Berkeley National Lab, 
       Berkeley, CA 94720} 
 
\maketitle 
 
\begin{abstract} 
  The  high $p_{\rm T}>3$~GeV azimuthal 
  asymmetry, $v_2(p_{\rm T})$, in non-central nuclear collisions at RHIC    
  is shown to be a sensitive measure of the initial parton density  
  distribution of the produced quark-gluon plasma.  
  A generalization of the Gyulassy-Levai-Vitev (GLV) non-abelian  
  energy loss formalism including  Bjorken 1+1D 
  expansion as well as important kinematic constraints is  used. 
\vspace{.2cm} 
 
\end{abstract} 
 
\begin{multicols}{2}  
   
{\em Introduction.\,}  
In order to interpret data on nuclear collisions 
from recent Relativistic Heavy Ion Collider (RHIC)  
experiments~\cite{phobos,starv2,phenix}, it is  obviously necessary to  
have knowledge of the {\em initial conditions}. 
Currently, there is an order of magnitude uncertainty 
in the initial produced gluon density,  
$\rho_g(\tau_0)\sim 10-100/{\rm fm}^3$, in central $Au+Au$ at  
$\sqrt{s}=130$~AGeV since widely different models~\cite{hijing,ekrt} 
are consistent~\cite{xnmg00}  with PHOBOS data~\cite{phobos}. 
We note that recent PHENIX data~\cite{phenix} appear to be 
inconsistent with at least  
one class (final state~\cite{ekrt})  of gluon saturation models. 
It is  essential, however, to check  this  
with other observables  as well. 
High $p_{\rm T}$ observables are ideally suited for this task 
because they provide a measure~\cite{hijing}   
of the total energy loss,  $\Delta E$,  of fast partons, 
resulting from medium induced  non-abelian  radiation along their  
path~\cite{mgxw,bdmsexp}. 
For intermediate jet energies ($E<20$~GeV), the predicted~\cite{glv2,levai} 
gluon energy loss in a {\em static} plasma  
of density $\rho_g$ and thickness, $L$ is approximately 
$\Delta E_{GLV}\sim E(L/6\;{\rm fm})^2\rho_g/(10/{\rm fm}^3)$. 
The approximate linear dependence of $\Delta E$  on $\rho_g$   
is the key that enables high $p_{\rm T}$ observables  
to convey information about the initial conditions. However,  
$\Delta E$ also depends non-linearly on the geometry, $L$,  
of the plasma and therefore differential observables which have well  
controlled geometric dependences are also highly desirable. 

A new way to probe $\Delta E$ in variable geometries 
was recently  proposed  in Ref.~\cite{wangv2}.  
The idea is to exploit the  spatial azimuthal asymmetry of non-central  
nuclear collisions.  
The dependence of $\Delta E$ on the path length $L(\phi)$  
naturally results in a pattern of azimuthal asymmetry of  high  
$p_{\rm T}$ hadrons which  can be measured  via the differential elliptic  
flow parameter (second Fourier coefficient),  
$v_2(p_{\rm T})$~\cite{starv2}. In this letter, we predict  
$v_2(p_{\rm T}>2\;{\rm GeV})$ for two  models of  initial  
conditions~\cite{xnmg00} which differ by an order of magnitude. 
We first generalize the finite energy  GLV theory~\cite{glv2} 
to take into account the  expansion (neglected in~\cite{levai,wangv2}) 
of the produced gluon-dominated plasma  
 while retaining kinematic  
constraints important for intermediate jet energies. 
Another novel element of the analysis is a discussion of the  
interplay between the azimuthally asymmetric soft  
(hydrodynamic~\cite{{Kolb:2000fh}}) and hard (quenched 
jet) components of the final hadron distributions. We show that  
the combined pattern of jet quenching  in the single inclusive 
spectra and the differential elliptic flow  at high $p_{\rm T}$ provide  
complementary tools that can determine the density as well as the  
spatial distribution of the quark-gluon plasma  
created  at RHIC. 
 
{\em  Hadron transverse momentum distributions.\,}  
It is useful to  
decompose the nuclear geometry dependence of  invariant hadron  
distributions produced in $A+B\rightarrow h+X$ at  impact parameter 
${\bf b}$ into a phenomenological ``soft'' and perturbative  
QCD  (pQCD) calculable ``hard'' components as 
\begin{eqnarray} 
{dN_{AB}({\bf b})  } &=&\;  
N_{part}({\bf b})\, {dN_s({\bf b}) } 
+ T_{AB}({\bf b}) \, {d\sigma_h ({\bf b})} \; , 
\label{decomp} 
\end{eqnarray} 
where $N_{part}({\bf b})$ is the number of nucleon participants, 
and $T_{AB}({\bf b})=\int  d^2{\bf r} \;T_A({\bf r})T_B({\bf r}-{\bf b})$ 
is the Glauber profile density per unity area in terms 
of  nuclear thickness functions, $T_A({\bf r})=\int dz \;\rho_A({\bf r},z)$.  
The computable lowest order pQCD differential cross section for  
inclusive $p+p\rightarrow h+X$ production is  given by 
\begin{eqnarray} 
E_{h}\frac{d\sigma_h^{pp}}{d^3p} &=&\! 
K   \sum_{abcd}\! \int\! \!dx_a  
dx_b f_{a/p}(x_a,Q^2_a) f_{b/p}(x_b,Q^2_b) \nonumber \\ 
&\;&\frac{d\sigma}{d{\hat t}}(ab\rightarrow cd) 
 \frac{D_{h/c}(z_c,{Q}_c^2)}{\pi z_c} \; , 
\label{hcrossec} 
\end{eqnarray} 
where $x_a, x_b$ are the initial parton momentum fractions,  
$z_c=p_h/p_c$ is the final hadron momentum fraction, 
$f_{a/p}(x_a,Q^2_a)$ are the 
parton distribution functions and  
$D_{h/c}(z_c,{Q}_c^2)$ is the fragmentation function for 
$c\rightarrow h$. 
The UA1 data on $p\bar{p}$ hadron  
production with $p_{\rm T} > 1$~GeV can be  
well reproduced with the above formula using  
$Q^2=p_{\rm T}^2/2$, $K=2$ and  
Martin-Roberts-Sterling~\cite{mrsd} (MRSD-') structure functions.
 
In nuclear collisions jet quenching can modify the hard cross section 
by changing the kinematic variables of the effective fragmentation 
function. We follow Ref.~\cite{wangv2} and include this effect 
by replacing the vacuum fragmentation function in  
Eq.~(\ref{hcrossec}) by an effective quenched one 
\begin{eqnarray} 
&&z_c D_{h/c}^\prime (z_c,Q^2_c) = z_c^\prime D_{h/c} 
(z_c^\prime,Q^2_{c^\prime})  +N_g z_g D_{h/g}(z_g,Q^2_g)\;, \nonumber \\ 
&&z_c^\prime = \frac{p_h}{p_c-\Delta E_c(p_c,\phi)} \;, 
\quad z_g  =  \frac{p_h}{\Delta E_c(p_c,\phi)/N_g} \; , 
\label{modfrag} 
\end{eqnarray} 
where $z_c^\prime,z_g$ are the rescaled momentum fractions. 
The first term is the fragmentation function of 
the jet $c$ after losing  energy $\Delta E_c(p_c,\phi)$ 
due to  medium {\em induced} gluon radiation. 
The second term is the  feedback due to the fragmentation  
of the $N_g(p_c,\phi)$ radiated gluons.  
The modified fragmentation function satisfies the  
sum rule $\int dz_c\; z_c D_{h/c}^\prime (z_c,Q^2_c)=1$.  
 
{\em Energy loss in a longitudinally expanding plasma.\,} 
The GLV reaction operator formalism~\cite{glv2} expands 
the  radiative energy loss formally in powers of the mean  
number, $\chi$, of interactions that the jet of energy $E$  
suffers along its path of propagation through dense matter. 
For a jet produced at point $\vx_0$, at time $\tau_0$, 
in an {\em expanding}  and possibly azimuthally asymmetric 
gluon plasma of density $\rho(\vx,\tau)$, 
the opacity in direction $\vh(\phi)$ is  
\begin{equation} 
\chi(\phi) =\int_{\tau_0}^\infty d\tau \;  
\sigma(\tau)\rho(\vx_0 + \vh(\phi)(\tau-\tau_0),\tau)\;. 
\label{opac} 
\end{equation} 
Note that  the gluon-gluon elastic cross section,   
$\sigma(\tau)=9\pi \alpha_s^2/2\mu^2_{eff}(\tau)$,  and 
the density may  vary along the path.  
For a finite jet energy, $E$, the approximate 
upper kinematic bound of medium induced momentum transfers is  
$|{\bf q}(\tau)_{\max} | \approx \sqrt{3  \mu(\tau) E }$ 
and $\mu_{eff}^2(\tau)=\mu^2(\tau)(1+\mu^2(\tau)/{\bf q}^2(\tau)_{\max} )$. 
The explicit closed form expression for the $n^{\rm th}$ order 
opacity expansion  of the gluon radiation double differential distribution 
for a static medium is given in Ref.~\cite{glv2}. Fortunately, 
 the opacity expansion converges very rapidly due to the formation 
time physics, and  the first order term was found  to give the dominant  
contribution. Higher order  corrections decrease rapidly with energy. 
All numerical results in this letter include  2$^{\rm nd}$ and  
3$^{\rm rd}$ order correction factors  computed in the static plasma  
limit~\cite{glv2}. 
We also include  finite kinematic bounds on the  
transverse momentum, ${\bf k}_{\max}^2=\min\, [4E^2x^2,4E^2x(1-x)]$ 
and ${\bf k}_{\min}^2=\mu^2$,  for  gluons with  
light-cone  momentum fraction $x$. Finite   
kinematics  reduces energy loss at intermediate jet  
energies~\cite{glv2,levai} 
 as compared to  
the asymptotic formalism~\cite{bdmsexp}.

The dominant (generalized) first order radiation {\em intensity}   
distribution~\cite{glv2}  that holds  also for expanding plasmas  
is given by ($z=\tau$) 
\beqar 
\frac{dI^{(1)}}{dx} &=&  
\frac{9 C_R   E}{\pi^2}  
\int\limits^\infty_{z_0} \!d z\,\rho(z)\!\!\! 
\int\limits^{|{\bf k}|_{\rm max} } \!\!\! d^2{\bf k}\, \alpha_s\!\!\! 
\int\limits^{|{\bf q}|_{\rm max} }  \!\!\! 
\frac{d^2{\bf q}\; \alpha_s^2}{({\bf q}^2+\mu(z)^2)^2} \, 
\nonumber \\[0.ex] 
&\;& \hspace{-0.0in} 
\frac{{\bf k}\cdot{\bf q}}{{\bf k}^2({\bf k}-{\bf q})^2}  
\left[ 1 
-   \cos \left (\,\frac{({\bf k}-{\bf q})^2}{2 x E}(z-z_0) \right) 
\right] \; . 
\eeqar{ndif1}  
In order to compare to previous asymptotic results~\cite{bdmsexp} 
for  expanding plasmas, consider a density of the form  
\beq 
\rho(z)=\rho_0\left(\frac{z_0}{z}\right)^\alpha \theta(L-z)\;, 
\eeq{rhoex} 
where $\alpha=0$ corresponds to a static uniform medium of thickness $L$, 
while $\alpha=1$ to a more realistic Bjorken 1+1D expansion of the plasma 
(transverse to the jet propagation axis).   
Analytic expressions can be obtained  only for asymptotic jet energies 
when the kinematic boundaries can be ignored~\cite{bdmsexp}.  
If we  set ${\bf q}^2_{\max} = {\bf k}^2_{\max} = \infty$, neglect 
the running $\alpha_s$ and change  
variables ${\bf k}-{\bf q}\rightarrow {\bf k}$, 
$u={\bf k}^2/\mu^2(z)$ and $w={\bf q}^2/\mu^2(z)$,  
then Eq.~(\ref{ndif1}) reduces to 
\beqar 
\frac{dI^{(1)}}{dx} &=&E\, \frac{2C_R\alpha_s}{\pi}  
\int^\infty_{z_0} d z\,\sigma(z)\rho(z)\; 
f(Z(x,z))  
\;, 
\eeqar{ndif11}  
where 
$ 
Z(x,z)=(z-z_0)\mu^2(z)/2 x E$ and 
\beqar 
f(x,z)&=& \int_0^\infty \frac{du}{u(1+u)}   
\left[ 1 -   \cos \left (\,u Z(x,z) \right) 
\right]  \nonumber   \\ 
&\approx&  \frac{\pi Z}{2}+\frac{Z^2}{2}\log(Z) + {\cal O}(Z^2) \;. 
\eeqar{fz} 
For a target of thickness $L$, the small $Z(x,z)$ limit applies as long as 
$x \gg x_c= L\mu^2(L)/2 E$. In that domain $dI/dx\propto 1/x$. 
For $x \ll x_c$, $f(Z)\approx \log \, Z$ and $dI/dx \propto \log\, 1/x$ 
is integrable to $x=0$. 
 
By integrating over $x$, the total energy loss is 
\beqar 
\Delta E &=&  
  E \frac{2C_R\alpha_s}{\pi} \int_0^1 dx  
\int^\infty_{z_0} d z\,\sigma(z)\rho(z)\; 
f(Z(x,z))\nonumber \\[.5ex] 
&\approx& \frac{C_R\alpha_s}{2}\int^\infty_{z_0} d z\, 
\frac{\mu^2(z)}{\lambda(z)}\;(z-z_0) \log \frac{E}{\mu(z)}\; , 
\eeqar{de11}  
which is an approximately linearly weighed line integral 
over the local transport 
coefficient $(\mu^2(z)/\lambda(z))\log E/\mu(z)=  
9 \pi\alpha_s^2 \rho(z)  \tilde{v} / 2 $.  
For a uniform and expanding plasma as in~(\ref{rhoex}) 
\beqar 
\Delta E_\alpha(L,z_0)  
&\approx& \frac{C_R\alpha_s}{2}  
\left(\frac{\mu^2(z_0)z_0^\alpha}{\lambda(z_0)}\right) 
\left(\frac{L^{2-\alpha}-z_0^{2-\alpha}}{2-\alpha}\right)\,\tilde{v} 
\nonumber \\ 
&=&\frac{C_R\alpha_s}{2} \frac{\mu^2(L)L^\alpha}{\lambda(L)}\; 
\frac{L^{2-\alpha}}{2-\alpha}\,\tilde{v} \;. 
\eeqar{deexp} 
Here $\tilde{v}=\log E/\mu$ and  
we used that $\mu^2(L)L^\alpha/\lambda(L)$ is 
a constant independent of $L$ for this type of expansion 
and took the $z_0\rightarrow 0$ limit. We therefore recover the 
asymptotic Baier-Dokshitzer-Mueller-Schiff (BDMS) and Zakharov (Z) energy loss
for both static  and  expanding media~\cite{bdmsexp}. 
We note that for Bjorken expansion, the asymptotic  
energy loss can be expressed 
in terms of the initial gluon rapidity density as 
\beqar 
\Delta E_{\alpha=1}(L)&=& \ 
\frac{9C_R\pi\alpha_s^3}{4} \left(\frac{1}{\pi R^2}  
\frac{dN^g}{dy} \right) \, L\,\log \frac{E}{\mu} \;.  
\eeqar{debj} 
If we vary $L=R\propto A^{1/3}$ by varying the nuclear size,  
then nonlinearity in $L$ arises because 
$dN^g/dy \propto A^{1+\delta}$. For  HIJING initial conditions~\cite{hijing}   
$\delta=1/3$,  while in the EKRT saturation model~\cite{ekrt} 
$\delta \approx 0$. 
 
{\em Implications of nuclear geometry.} 
For nucleus-nucleus collisions  
the co-moving  plasma  produced in an $A+B$ reaction at impact  
parameter ${\bf b}$ at formation time  
$\tau=z_0$ has a transverse 
coordinate distribution given by  
\beq 
\rho_g({\bf r},z=0,\tau=z_0)= \frac{1}{z_0} 
\frac{d\sigma^{\rm jet}}{dy} \,T_A({\bf r}) 
\,T_B({\bf r} -{\bf b})\;, 
\eeq{rhog} 
where $d\sigma^{\rm jet}/dy$ is the pQCD mini-jet cross section  
in $pp$ collisions at a given $\sqrt{s}$. 
Note that taking into account also the 
2D transverse expansion  
causes the density to decrease 
somewhat   
faster than Eq.~(\ref{rhoex}).  
However, we  found numerically that  
transverse expansion can be ignored  
in the first approximation. 
\begin{center} 
\vspace*{8.9cm} 
\includegraphics{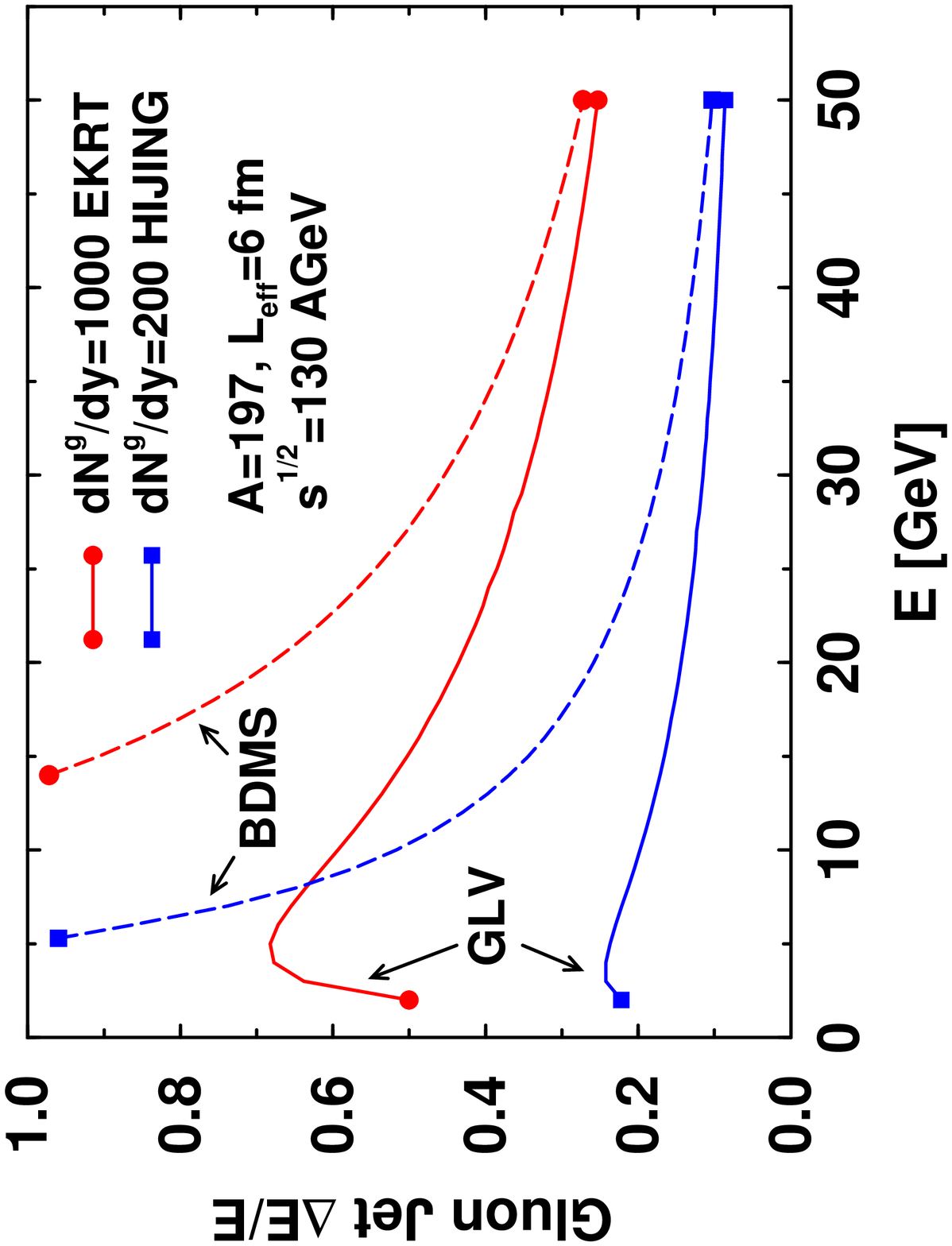} 
\vspace{-2.7cm} 
\end{center} 
\begin{center} 
\begin{minipage}[t]{8.5cm} 
         { FIG. 1.} {\small  The  
GLV fractional  
energy loss  
 in 
Bjorken expanding 
gluon plasma 
with ~\cite{hijing,ekrt} $dN^g/dy \simeq 200, 1000$.  
} 
\end{minipage} 
\end{center} 
\begin{center} 
\vspace*{8.7cm} 
\includegraphics{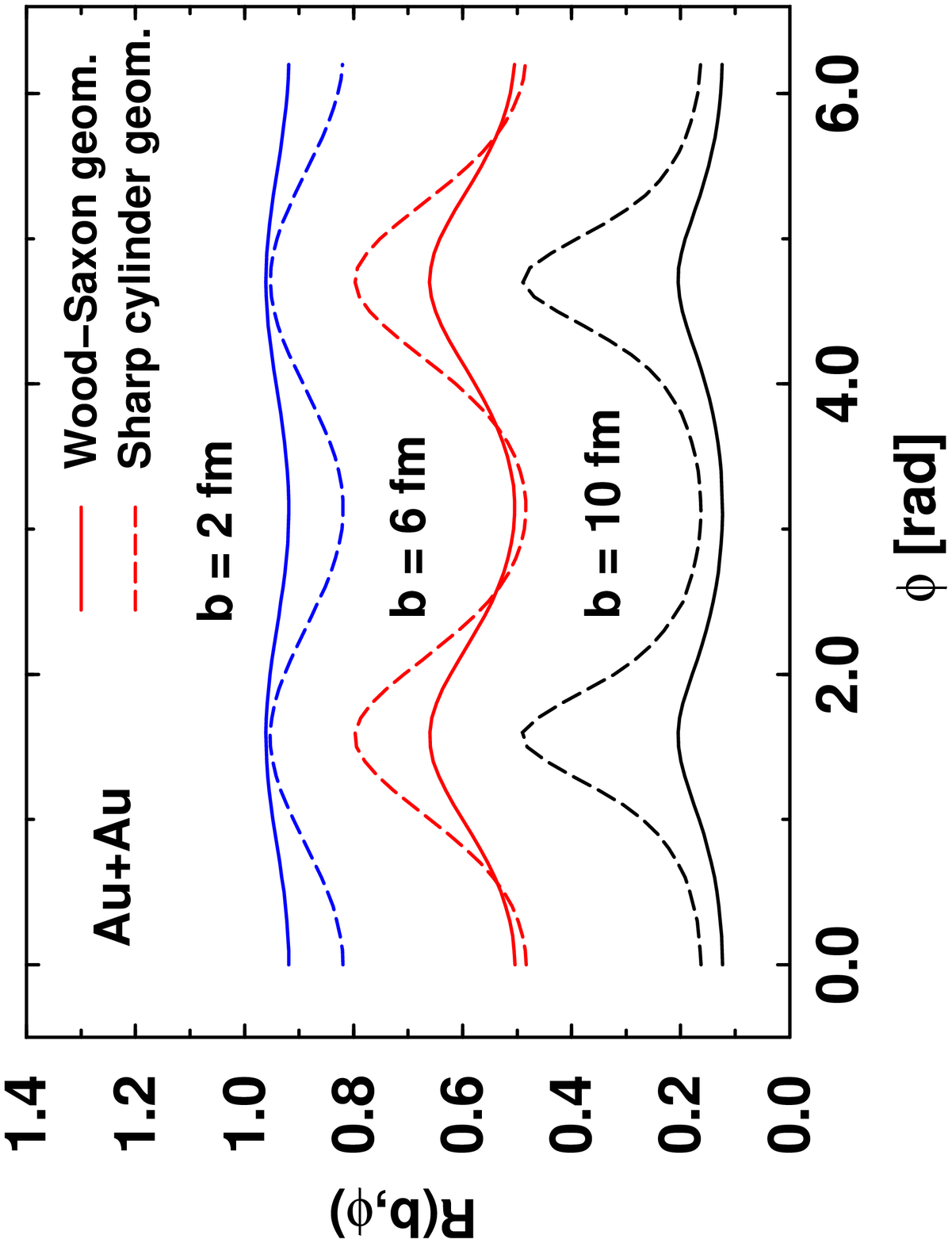} 
\vspace{-2.7cm} 
\end{center} 
\begin{center} 
\begin{minipage}[t]{8.7cm}   
      { FIG. 2.} {\small The modulation function  $R({\bf b},\phi)$    
is plotted vs. $\phi$ for several impact parameters.
and Wood-Saxon vs 
cylinder geometries. 
}  
\end{minipage} 
\vspace{.5cm}
\end{center} 
In the linear $f(Z)\approx \pi Z/2$~(\ref{fz}) and  
Bjorken  approximations, 
the total energy loss is proportional to the line  
integral~(\ref{ndif11},\ref{de11}) along the jet trajectory 
${\bf r}(z,\phi)={\bf r}+\hat{v}(\phi)(z-z_0)$, 
averaged over the distribution  
of the jet production points  
\begin{eqnarray} 
F({\bf b},\phi) &=&   
 \int d^2{\bf r} \; \frac{T_A({\bf r})T_B({\bf r}-{\bf b})} 
{T_{AB}({\bf b})}  
\int_{z_0}^\infty 
 dz \; z\; \left(\frac{z_0}{z}\right)^\alpha \nonumber \\ [1ex] 
&\;& \quad T_A({\bf r}(z,\phi)) T_B({\bf r}(z,\phi)-{\bf b}) \;, 
\label{linint} 
\end{eqnarray} 
where $T_A({\bf r}),T_B({\bf r- b})$ and   
$T_{AB}({\bf b})$ depend on the geometry. 
In particular, for a sharp uniform cylinder of radius $R_{\rm eff}$ 
 one readily gets ${T_A}({\bf r})=(A/\pi R^2_{\rm eff}) 
\theta (R_{\rm eff}-|{\bf r}|)$ and  
${T_{AB}}(0)=A^2/\pi R^2_{\rm eff}$. 
We can therefore define the  effective radius of the sharp cylinder  
equivalent to a diffuse Wood-Saxon geometry via 
 \begin{eqnarray} 
&& F ({\bf 0},\phi)_{\rm Wood-Saxon} =   
F ({\bf 0},\phi)_{\rm Sharp \;\; cylinder} \; . 
\label{linconstr} 
\end{eqnarray}  
For $Au+Au$ collisions and $\alpha=1$, 
 Eq.~(\ref{linconstr}) gives $R_{\rm eff}\approx 6$~fm.  
Eq.~(\ref{ndif1}) can then be integrated numerically to give  
$\Delta E(0)/E$, allowing $\alpha_s$ to run 
and including kinematical bounds.  
Fig.~1 illustrates the fractional energy loss  
for gluon jets at $b=0$ for a broad range of initial gluon   
densities~\cite{hijing,ekrt}. 
 
For a non-vanishing  impact parameter ${\bf b}$  
and jet direction ${\hat{v}(\phi)}$,  we calculate the energy loss as 
\begin{equation} 
\frac{\Delta E({\bf b},\phi )}{E} 
= \frac{F({\bf b},\phi )}{F({\bf 0},\phi)} \, \frac{\Delta E(0)}{E}  
\equiv  
R({\bf b},\phi ) \, \frac{\Delta E(0)}{E} \;, 
\label{separat} 
\end{equation} 
where the modulation function $ R({\bf b},\phi)$ captures in the  
{\em linearized} approximation the ${\bf b}$ and $\phi$  
dependence of the jet energy loss.  
Fig.~2 shows the   $R({\bf b},\phi)$  modulation factor 
plotted against the azimuthal angle $\phi$ for impact parameters  
${\bf b}=2, 6, 10$~fm. Note that $R({\bf b},\phi)$ reflects not only  
the dimensions of the characteristic ``almond" cross section  
shape of the interaction volume but also the  
rapidly decreasing  initial plasma density  as a function 
of the impact parameter. 
 
{\em Phenomenological soft ``hydrodynamic'' component.\,} 
In order to compare to the new STAR data~\cite{starv2} at $p_{\rm T}<2$~GeV, 
we must also take into account the  soft non-perturbative component 
that cannot be computed with the eikonal jet quenching formalism above. 
In~\cite{wangv2} this was simply modeled by an azimuthally {\em symmetric} 
exponential form. However, in non-central $A+B$ reactions  
the low $p_{\rm T}$ hadrons are also expected to exhibit azimuthal 
asymmetry caused by hydrodynamic like flow effects~\cite{Kolb:2000fh}. 
We  therefore model the low $p_{\rm T}$ soft component  
here with  the following ansatz: 
\begin{equation} 
\frac{dN_s ({\bf b})}{dyd^2{\bf p}_{\rm T}} = 
\frac{dn_s}{dy}\frac{e^{-p_{\rm T}/T_0}}{2\pi T^2_0} 
\left(1+2 v_{2s}(p_{\rm T})\cos(2\phi)\right) 
\; , 
\end{equation} 
where we take $T_0\approx 0.25$ GeV 
and incorporate the hydrodynamic elliptic flow predicted  
in~\cite{Kolb:2000fh} and  found to grow monotonically  
with $p_{\rm T}$ as   
\begin{equation} 
v_{2s}(p_{\rm T}) \approx {\rm tanh}(p_{\rm T}/(10\pm 2\;{\rm GeV})) 
\;\; .\end{equation} 
It is important to emphasize that  
hydrodynamic flow was found~\cite{Kolb:2000fh} 
to be less sensitive to the initial conditions 
than the high $p_T$  jet quenching reported here. 
 
With the inclusion of this non-perturbative soft component, 
it follows  from Eq.~(\ref{decomp}) that  
the effective differential flow  is 
\begin{center} 
\vspace*{8.5cm} 
\includegraphics{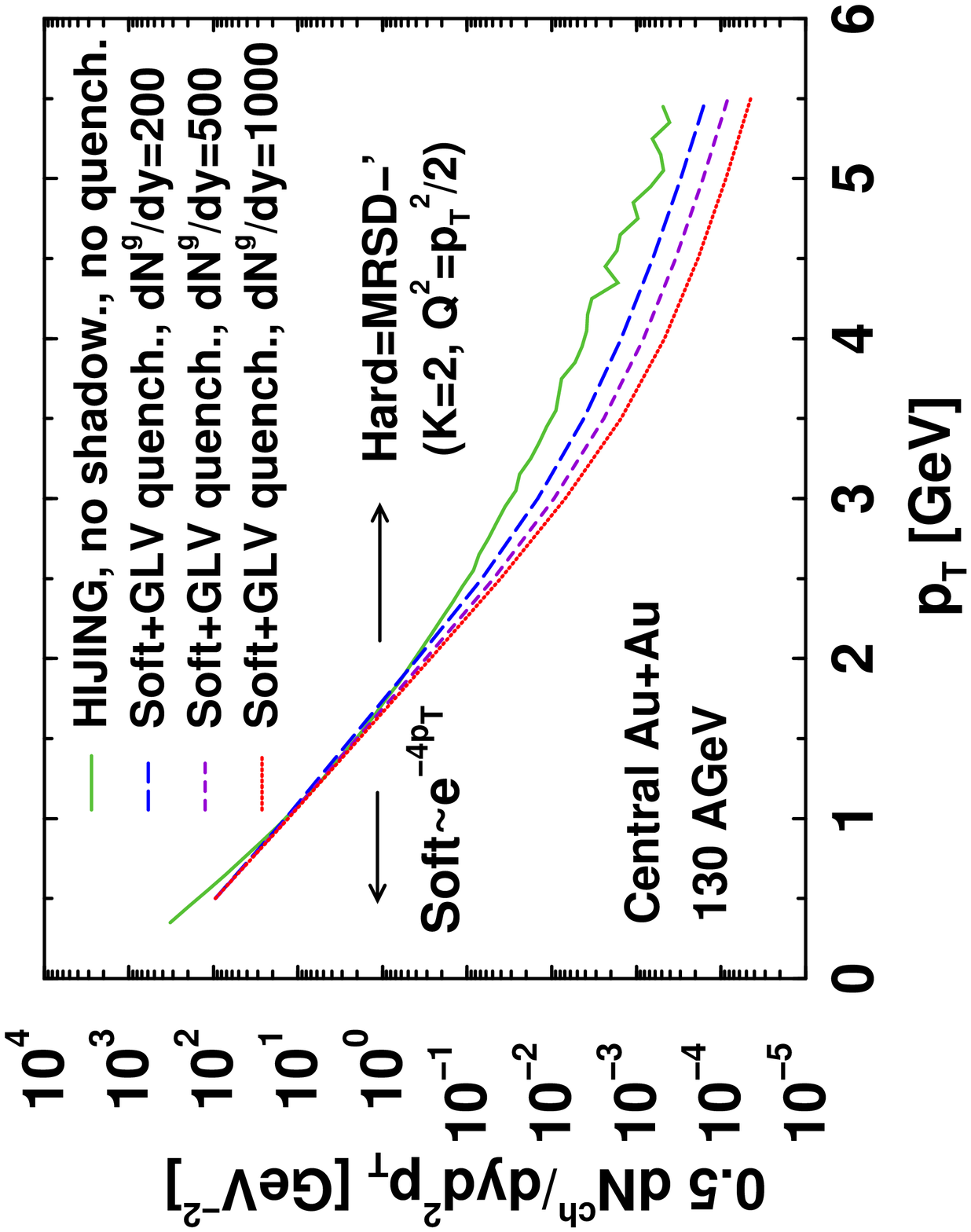} 
\vspace{-2.8cm} 
\end{center} 
\begin{center} 
\begin{minipage}[t]{8.6cm}   
   { FIG. 3.} {\small Sensitivity of central-collision 
inclusive hadron distributions to initial conditions and energy loss in  
the two component hydrodynamic + quenched jet model.   
}  
\end{minipage} 
\end{center} 
\begin{center} 
\vspace*{8.7cm} 
\includegraphics{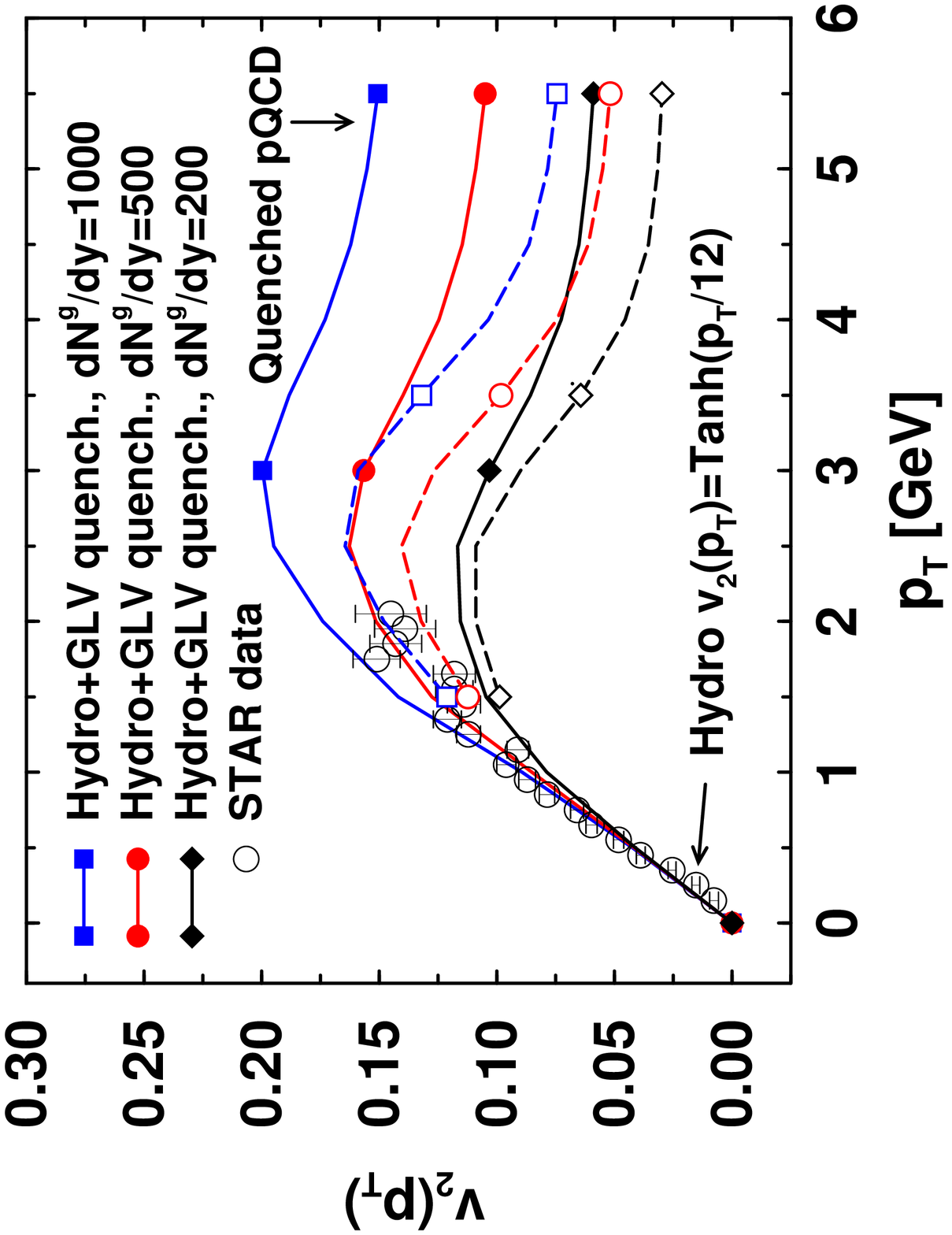} 
\vspace{-2.5cm} 
\end{center} 
\begin{center} 
\begin{minipage}[t]{8.6cm}   
   { FIG. 4.} {\small The interpolation of $v_2(p_{\rm T})$ 
between the soft hydrodynamic~\protect{\cite{Kolb:2000fh}} and 
hard pQCD regimes is shown for $b=7$ fm. 
Solid (dashed) curves  
correspond to 
cylindrical (Wood-Saxon) 
geometries. 
} 
\end{minipage} 
\end{center} 
\begin{equation} 
v_{2}(p_{\rm T}) \approx  
\frac{v_{2s}(p_{\rm T}) dN_s + v_{2h}(p_{\rm T}) dN_h} 
{ dN_s+dN_h } \;. 
\label{v2interp} 
\end{equation} 
This interpolates between the hydrodynamic and the pQCD regimes because 
at high $p_{\rm T}$, $dN_h\gg dN_s$. For our numerical estimates 
the low $p_{\rm T}$ interpolation is achieved 
by multiplying the pQCD curves with a switch function  
$[1+{\rm tanh}(3 (p_{\rm T}-1.5\,{\rm GeV}))  ]/2 $. 
 
{\em  Conclusions.\,} 
Fig.~3 shows the inclusive charged particle transverse 
momentum distribution in central $Au+Au$ collisions 
with three models of initial gluon densities $dN^g/dy=1000, 500, 200$.  
We see that jet quenching can be disentangled from the soft  
hydrodynamic  component 
only for transverse momenta $p_{\rm T}>4$~GeV.  
In that high $p_{\rm T}$ region 
there is an approximately constant suppression 
 relative to  the unquenched 
(HIJING) distribution due to the approximately {\em linear} energy  
dependence~\cite{levai} of 
the GLV energy loss~\cite{glv2}. The suppression increases systematically 
with increasing initial plasma density and thus  provides 
an  important constraint on the maximum initial parton densities produced 
in ${\bf b}=0$ collisions. 
 
Fig.~4 shows the predicted pattern of high $p_{\rm T}$ anisotropy. 
Note the difference between sharp cylinder and diffuse Wood-Saxon geometries 
at ${\bf b}=7$~fm, the 
characteristic impact parameter of minimum bias events. While  
 the central (${\bf b} = 0$)  
inclusive quenching is insensitive to the 
density profile (due to Eq.~(\ref{linconstr})), 
 non-central events clearly exhibit  large sensitivity 
to the actual distribution. We checked numerically that transverse 
expansion with $v_\perp = 0.5c$ can be ignored since  
it reduces the jet quenching effects in Figs.~3,4  
by $<$20\% at high $p_{\rm T}$. 
 
We conclude that $v_2(p_{\rm T}>2\;{\rm GeV},{\bf b})$  
provides essential complementary information about the geometry and impact 
parameter dependence of the 
initial conditions in $A+A$. In particular, the rate at which the $v_2$ coefficient  
decreases at high $p_{\rm T}$ is an  indicator of the 
diffuseness of that  geometry.  
 
We thank P. Huovinen,  
R. Snellings, A. Poskanzer, and H.J. Ritter for  
stimulating discussions. 
This work was supported by the  
the U.S. DOE under  
DE-AC03-76SF00098 and 
DE-FG-02-93ER-40764 and by the NSFC under 
No. 19928511.  

\vfill\eject 
\end{multicols} 
\end{document}